\documentstyle[12pt]{article}
\begin{document}

\title {\bf A Statistical Measure of Complexity}
\author{R. L\'{o}pez-Ruiz$^{1,\ddag}$, H.L. Mancini$^2$ and X. Calbet$^3$ \\       
                                   \\
{\small 
$^1$ Laboratoire de Physique Statistique, ENS, 24 rue de Lhomond, Paris (France)} \\
{\small 
$^2$ Departamento de F\'{\i}sica, Universidad Privada de Navarra, Pamplona (Spain)} \\
{\small
$^3$ Instituto de Astrof\'{\i}sica de Canarias, La Laguna, Tenerife (Spain)}  
\date{ }}

\maketitle
\baselineskip 6mm
 
\begin{center} {\bf Abstract} \end{center}
 
A measure of {\it complexity} based on a probabilistic description 
of physical systems is proposed. This measure incorporates the main features
of the intuitive notion of such a magnitude. It can be
applied to many physical situations and to different
descriptions of a given system. Moreover, the calculation of its value does not
require a considerable computational effort in many cases of physical interest.

PACS number(s): 05.20.-y, 02.50.+s, 02.70.+d 

\newpage
 
On the most basic grounds, an object, a
procedure, or system is  said to be "complex" when it does not match 
patterns regarded as simple. This sounds rather like an oxymoron 
but common knowledge tells us what is simple and complex: 
simplified systems or idealizations are always a starting point to solve
scientific problems. The notion of "complexity"  
in physics  \cite{parisi,anderson}  starts
 by considering the perfect crystal and the 
isolated ideal gas as examples of simple models and therefore as systems 
with zero "complexity".
Let us briefly recall their main characteristics with
"order", "information" and "equilibrium". \par

A perfect crystal is completely ordered and the  
atoms are arranged  following stringent rules of symmetry. 
The probability distribution for the states accessible to the perfect
crystal is centered around a prevailing state of
perfect symmetry. A small piece of "information" is enough to describe
the perfect crystal: the distances and 
the symmetries that define the elementary cell.
The "information"  stored in this system can be considered minimal.
On the other hand, the isolated ideal gas is completely
disordered. The system  can be found in any of its accessible states with 
the same probability. All of them contribute in equal measure to
the "information" stored in the ideal gas.
It has therefore a maximum "information". These two  simple systems
are extrema in the scale of "order" and "information". It follows that
the definition of "complexity" must not be made in terms of just "order"
or "information". \par 
It might seem reasonable to propose a measure of "complexity"
by adopting some kind of distance from the equiprobable distribution of the
accessible states of the system. Defined in this way,
"disequilibrium"  would give an idea of
the probabilistic hierarchy of the system. 
"Disequilibrium" would be different from zero if there are privileged,
or more probable, states among those accessible. But this would not work.
Going back to the two examples we began with, it is readily seen that a perfect
crystal is far from an equidistribution among the accessible states
because one of them is totally prevailing, and so "disequilibrium"
would be maximum. For the ideal gas, "disequilibrium" would be zero by
construction. Therefore such a distance or "disequilibrium" (a measure
of a probabilistic hierarchy) cannot be directly associated 
with "complexity". \par
In figure 1. we sketch an intuitive qualitative behaviour for "information"
$H$ and "disequilibrium" $D$ for systems ranging from 
the perfect crystal to the ideal gas. This graph suggests that
the product of these two quantities  could be used 
as a measure of "complexity": $C = H \cdot D$ (Fig. 1.).
The function $C$ has indeed the features and asyntotical properties
that one would expect intuitively: it vanishes for the 
perfect crystal and for the
isolated ideal gas, and it is different from zero for the rest of the
systems of particles. We will follow these guidelines to establish a quantitative
measure of "complexity". \par
Before attempting any further progress, however, we must recall that 
"complexity" cannot be measured univocally, because it depends on the
nature of the description (which always involves a reductionist process)
and  on the scale of observation. Let us take an example to illustrate
this point. A computer chip can look very different at different scales.
It is an entangled array of electronic elements at microscopic scale but 
only an ordered set of pins attached to a black box at a macroscopic scale.\par

We shall now discuss a measure of "complexity" based on the 
statistical description of systems.
Let us assume that the system has N accessible states $\{x_1,x_2,...,x_N\}$ 
when observed at a given scale. We will call this an N-system. 
Our understanding of the behaviour of this system determines the 
corresponding probabilities $\{p_1,p_2,...,p_N\}$ (with the 
condition $\sum_{i=1}^{N}p_i =1$) of each state ($p_i\neq 0$ for all $i$). 
Then the knowledge of the underlying
physical laws at this scale is incorporated into a probability distribution 
for the accessible states. It is possible to find a quantity
measuring the amount of "information".
Under to the most elementary conditions of 
consistency, Shannon \cite{shannon} determined the unique function 
$H(p_1,p_2,...,p_N)$ that accounts for the "information" stored
in a system:
\begin{equation}
H = -K \sum_{i=1}^{N} p_i\log p_i 
\end{equation}
where $K$ is a positive constant. The  quantity $H$ is called 
{\it information}. In the case of 
a crystal, a state $x_c$ would be the most probable
$p_c\sim 1$, and all others $x_i$ 
would be very improbable, $p_i\sim 0$ $i\neq c$. Then $H_c \sim 0$. 
On the other side, equiprobability
characterizes an isolated ideal gas, $p_i\sim 1/N$ so $H_g\sim K\log N$,
i.e., the maximum of information for a N-system.
(Notice that if one assumes equiprobability and $K=\kappa\equiv Boltzmann$ 
$constant$, $H$ is identified  with the thermodinamic
entropy ($S=\kappa\log N$)). Any other N-system will have an amount of
information between those two extrema. \par
Let us propose a definition of {\it disequilibrium} $D$ \cite{prigo}
in a N-system. The intuitive notion suggests that some kind of 
distance from an equiprobable distribution should be adopted.
Two requirements are imposed on the magnitude of $D$: $D>0$ in order to have a
positive measure of "complexity" and $D=0$ on the limit of equiprobability.
The straightforward solution is to add the quadratic distances of 
each state to the equiprobability  as follows: 
\begin{equation}
D = \sum_{i=1}^{N}(p_i - \frac{1}{N})^2
\end{equation}
According to this definition, a crystal has maximum disequilibrium
(for the dominant state,
$p_c\sim 1$, and $D_c\rightarrow 1$ for $N\rightarrow \infty$)
while the disequilibrium for an 
ideal gas vanishes ($D_g\sim 0$) by construction. For any other system 
$D$ will have a value between these two extrema. \par
We now introduce the definition of {\it complexity} $C$ of a $N$-system.
This is simply the interplay between the information stored in 
the system and its disequilibrium:
\begin{equation}
C = H \cdot D = -\left ( K\sum_{i=1}^{N} p_i\log p_i \right ) \cdot
\left (\sum_{i=1}^{N}(p_i - \frac{1}{N})^2 \right )
\end{equation} 
This definition fits the intuitive arguments.
For a crystal, disequilibrium is large but the information stored
is vanishingly small, so $C\sim 0$.
On the other hand, $H$ is large for an ideal gas, but $D$ is small, 
so $C\sim 0$ as well. Any other system will have an 
intermediate behavior and therefore $C>0$. \par
As was intuitively suggested, the definition of complexity (3) 
also depends on the {\it scale}.
At each scale of observation a new set of accessible states appears
with its corresponding probability distribution so that
complexity changes. Physical laws at each level
of observation allow us to infer the probability
distribution of the new set of
accessible states, and therefore different values for $H$, $D$ and $C$ 
will be obtained. (In the most complicated situations, where
there exist extremely many different states, there are methods 
to calculate functions of the probability distribution \cite{pochel}).
 The passage to the case of a continuum number of states,
$x$, is straighforward. Thus we must treat with probability distributions
with a continuum support, $p(x)$, and normalization condition
$\int_{-\infty}^{+\infty}p(x)dx=1$. Disequilibrium has the limit
$D=\int_{-\infty}^{+\infty}p^2(x)dx$ and the complexity is defined by:
\begin{equation} 
C=H\cdot D=-\left(K\int_{-\infty}^{+\infty}p(x)\log p(x)dx \right)
\cdot\left(\int_{-\infty}^{+\infty}p^2(x)dx \right)
\end{equation} \par

Direct simulations of the definition give the values
of $C$ for general N-systems.
The set of all the possible distributions $\{p_1,p_2,...,p_N\}$ 
where an N-system could be found 
is sampled. For the sake of simplicity $H$ is normalized to the interval
$[0,1]$. This magnitude is called $\overline{H}$.
Thus $\overline{H}=\sum_{i=1}^{N} p_i\log p_i/\log N$.
 For each distribution $\{p_i\}$ the normalized information 
 $\overline{H}(\{p_i\})$, and the disequilibrium 
$D(\{p_i\})$ (eq. 2) are calculated.
In each case the normalized complexity
$\overline{C}=\overline{H}\cdot D$ is obtained and the
pair $(\overline{H},\overline{C})$ stored.
These two magnitudes are plotted on a diagram $(\overline{H},
\overline{C}(\overline{H}))$ in order
to verify the qualitative behavior predicted in figure 1. For N=2
an analytical expression for the curve 
$\overline{C}(\overline{H})$ is obtained.
If the probability of one state is $p_1 =x$, that 
of the second one is simply $p_2 =1-x$. 
The complexity of the system will be:
\begin{equation}
\overline{C}(x)= \overline{H}(x)\cdot D(x)=-\frac{1}{\log 2}
[x\log\left(\frac{x}{1-x}\right)+ \log(1-x)]\cdot 2(x-\frac{1}{2})^2
\end{equation}
Complexity as a function of $\overline{H}$ is shown in
figure 2a. It vanishes for the two simplest 2-systems:
the crystal ($\overline{H}=0$; $p_1 =1$, $p_2 =0$) and the ideal gas 
($\overline{H}=1$; $p_1 =1/2$, $p_2 =1/2$). 
Let us notice that this curve is the simplest one that fulfills all
the conditions discussed in the introduction. 
The largest complexity is reached for $\overline{H}\sim 1/2$ and its value
is: $C(x\sim 0.11)\sim \overline{C}\cdot \log 2\sim 0.105$. 
For N$>$2 the relationship between 
$\overline{H}$ and $\overline{C}$ is not
univocal anymore. Many different distributions $\{p_i\}$ store
the same information $\overline{H}$ but have different complexity 
$\overline{C}$. Figure 2b. displays such a behavior for $N=3$. 
If we take the maximum 
complexity $\overline{C}_{max}(\overline{H})$ associated with each 
$\overline{H}$ a curve similar to the one shown in a 2-system is recovered.
Every 3-system will have a complexity below this line. In figure 2a.
curves $\overline{C}_{max}(\overline{H})$ for the cases 
$N=3,5,7$ are also shown. 
Let us observe the shift of the complexity-curve peak 
to smaller values of entropy for rising $N$. This fact 
agrees with the intuition telling us that the biggest complexity
(number of possibilities of 'complexification') be reached for lesser
entropies for the systems with bigger number of states.\newline
Also it is possible to compute these quantities in other relevant physical
situations and in continuum systems. \par

We can now go one a step further. The most important point is that the 
new definition should work in 
systems out of equilibrium. We use two examples of such systems 
where it is known that very complex dynamics could show up. They are the
logistic map and the 'Lorenz' map. \newline
{\bf Logistic Map}: The mapping 
$x_{n+1}=\alpha\cdot x_n (1-x_n)$, $\alpha\in [0,4]$,
is a well known chaotic system. There are two points in this system
where the behaviour is extremely "complex". The first is the accumulation point
of the subharmonic cascade. The second is the transition point from chaos to a
period three orbit via intermittency \cite{pomeau}. We are going to discuss
and study the complexity in the second case only, which
corresponds to parameter values around $\alpha_t \sim 3.8284$.
Complexity must increase
as $(\alpha -\alpha_t)\rightarrow 0$ 
when $\alpha <\alpha_t$ because the closer we get to the
critical point $\alpha_t$ the more improbable and unpredictable is
the firing and development 
of the intermittent bursts. On the contrary, when $\alpha >\alpha_t$
the system becomes periodic and does not have any complexity.
Complexity \cite{calculo} undergoes a rapid increase in the intermittency 
region and a sharp transition to zero at the transition point (Fig. 3a).
(This resembles the curve of the specific heat as a function of 
temperature in a second order phase transition). \newline
{\bf Lorenz Map}: Let us consider a mapping that mimics the behavior
and development of the Lorenz attractor. This is the simplest 
mapping that includes the main features of the first return
map of this attractor: \{$x_{n+1}=\alpha\cdot x_n$ if $x_n<0.5$ and 
$x_{n+1}=\alpha\cdot (x_n-1)+1$ if $x_n>0.5$\} where $\alpha\in (0,2)$. Its dynamic
evolution displays three different behaviors: 1) $R_1$: $\alpha\in (0,1)$. The
system goes to a fixed point ($0$ or $1$) and the output is a constant
signal; 2) $R_2$: $\alpha\in (1,2)$. A variable chaotic attractor is
present in all this region and a chaotic signal is obtained; 
3) $R_3$ : $\alpha\approx 2$. The Bernouilli-shift is reached and the output
signal is random.
Results for complexity are given in figure 3b. In region $R_1$,
$C$ vanishes because there is nothing to explain a constant signal.
In region $R_2$, $C\neq 0$ and shows a complicated
dependence on $\alpha$. Variations in $C$ are due to changes in
the structure of the underlying chaotic attractor. 
When $\alpha\rightarrow 2$ (region $R_3$) a random system is reached, 
and again $C=0$. \par

Let us return to the point at which we started the intitial
discussion. Any notion of complexity in physics \cite{parisi,anderson} should only
be made on the basis of a well defined or operational magnitude.
But two additional requirements are needed in order to obtain a 
good definition of complexity in physics: ($c1$) the new magnitude
must be measurable in many different physical systems and ($c2$) 
a comparative relationship and a physical interpretation 
between any two measurements should be possible. \par
Many different definitions of complexity have been proposed to date, mainly
in the realm of computational sciences. Among these, several can be cited: 
algorithmic complexity (Kolmogorov-Chaitin) \cite{kolmo,chaitin},
the Lempel-Ziv complexity \cite{lempel}, the logical depth of Bennett 
\cite{bennett}, the effective measure complexity of
Grassberger \cite{grass}, the complexity of a system based in its 
diversity \cite{huberman}, the thermodynamical depth \cite{lloyds},
complexities of formal grammars, etc.
The definition of complexity proposed in this work offers a new point of view,
based on a statistical
description of systems at a given {\it scale}. In this 
scheme the knowledge of the physical laws governing the dynamic evolution 
in that scale is used to find its accessible states and its
probability distribution. This process would inmediately indicate the
value of complexity. In essence this is nothing but an interplay
between the information stored by the system and the 
{\it distance from equipartition} (measure of a probabilistic hierarchy
between the observed parts) of the probability distribution of
its accessible states . 
Besides giving the main features of a "intuitive" notion
of complexity, we showed that it sucessfully enables us
to discern situations regarded as complex, both for a local
transition (Fig. 3a.) and for a global behavior (Fig. 3b.) in systems
out of equilibrium. 

{\bf Acknowledgements:}
We thank Prof. C. P\'{e}rez-Garc\'{\i}a from Universidad Privada de Navarra
for useful comments and references.
This research was funded by the DGICYT (Spanish Government) 
under grant PB-93-0708. One of us (R.L-R) thanks the French
Government for a research grant.


\newpage
\begin{center} Figure Captions \end{center}

{\bf Fig 1.} Sketch of the intuitive notion of the magnitudes of 
"information" (H) and "disequilibrium" (D) for the physical systems and
the behavior intuitively required for the magnitude 
"complexity". The quantity $C=H\cdot D$ is proposed to measure
such a magnitude. \par

{\bf Fig 2a.} Complexity ($\overline{C}=\overline{H}\cdot D$) as a function
of the information ($\overline{H}$) for a system with two accessible
states ($N=2$). Also curves of maximum complexity ($\overline{C}_{max}$) 
are shown for the cases: $N=3,5,7$. \par

{\bf Fig 2b.} In general, dependence of complexity ($\overline{C}$)
on information ($\overline{H}$) is not univocal: many distributions $\{p_i\}$
can present the same value of $\overline{H}$ but different $\overline{C}$.
This is shown in the case $N=3$. \par

{\bf Fig 3a.} Behavior of complexity $C$ in the transition point 
($\alpha_t\sim 3.8284$) where the system (logistic map) goes from a chaotic 
dynamics ($\alpha<\alpha_t$) to a period three 
orbit ($\alpha>\alpha_t$) by intermittency. \par

{\bf Fig 3b.} Complexity displayed by the 'Lorenz' map for a region
of the parameter space $\alpha\in [0,2]$. (See explanation in the text).
(Calculations have been done for a scale $n=12$, $n\equiv$ length 
of the binary strings analyzed).


\begin{thebibliography}{80}
\bibitem{calculo}
For each parameter value, $\alpha$, calculations of complexity have been 
done in the following steps:
(i) the dynamics is reduced to a binary sequence (its symbolic dynamics:
$0$ if $x<0.5$ and $1$ if $x>0.5$); (ii) the scale $n$ in these cases
is the length $n$ of the binary strings considered as states of the system.
Thus, $2^n$ states are possible in this case; (iii) the probability 
of each state is measured when the system is 
running over $2^{n+10}$ iterations, that is
binary sequences of $2^{n+10}$ digit-length; (iv) $H$, $D$ and $C$  
are calculated directly with that probability distribution.
\end{thebibliography}
\end{document}